# Emergent non-Hermitian contributions to the Ehrenfest and Hellmann-Feynman theorems

Georgios Konstantinou, Kyriakos Kyriakou, Konstantinos Moulopoulos

Department of Physics, University of Cyprus, PO Box 20537, 1678 Nicosia, Cyprus

**We point out that two of the most important theorems of Quantum Mechanics, the Ehrenfest theorem and the Hellmann-Feynman theorem, lack – in their standard form – important information: there are cases where non-Hermitian boundary contributions emerge. These contributions actually appear naturally, in order for the above theorems to be valid and applicable (i.e. in multiply-connected spaces), and this occurs for physical quantities that are not represented by well-defined self-adjoint operators (such as the position operator in a periodic potential, or in general Aharonov-Bohm configurations, either in real or in an arbitrary parameter-space, in the sense of Berry's adiabatic and cyclic procedures). In this short note, we report modifications of these two theorems when such non-Hermiticities appear, and we demonstrate how they resolve certain Quantum Mechanical paradoxes (most of them having been noticed in the past as violations of the so-called Hypervirial theorem in Quantum Chemistry). This resolution of paradoxes (essentially the re-establishment of applicability of the Ehrenfest theorem even in multiply-connected spaces) always proceeds through the appearance of certain generalized currents, in a theoretical picture with interesting structure (where a generalized continuity equation with a sink term shows up naturally).**

## I. INTRODUCTION

Well-known and fundamental theorems of Quantum Mechanics, such as the Ehrenfest and the Hellmann-Feynman theorems, are usually applied in the literature without considerations of their underlying limitations. And in the rare cases in which they *are* scrutinized (i.e. cases corresponding to operators that are not strictly self-adjoint), they are practically labeled as inapplicable (as i.e. in multiply-connected spaces). Simply put, we here show that in the latter cases we can still make use of the theorems, if we are willing to accept boundary terms that are usually thrown away after integrations; and we also show that these terms (a reflection of what could be viewed as emergent non-Hermiticity) may even hide important physical information. These theorems have historically played a major role in the formulation of Quantum Mechanics, the Ehrenfest theorem, for example, defining the 'velocity operator' as well as the 'force' operator, while the Hellmann-Feynman theorem being useful in also defining a velocity in the crystal momentum space, or, more generally, revealing information about the slopes of the energy bands in the Brillouin Zone. But is application of these theorems always as innocent as it is usually assumed? The answer seems to be in the negative. There are cases where additional boundary-related information has to be considered. This becomes necessary as the Hamiltonian operator itself might demonstrate hidden non-Hermiticity, leading to erroneous results (as i.e. in the Bloch crystal case, as we will see, where naive use of the Hellmann-Feynman theorem may lead to the erroneous conclusion that the slope of the energy bands must vanish!). Another example is an apparent additional boundary contribution to the standard velocity operator, that could transfer information between two systems through an interface. In this paper we magnify on such issues, and we show how these extra (non-Hermitian) boundary contributions actually correct (and resolve) previously noticed paradoxes regarding these theorems. It should be added that the non-Hermiticity discussed in the present article does not seem to have anything to do with the area of non-Hermitian Quantum Mechanics that has been developed in the last 2 decades after the seminal work of Bender and Boettcher[1]; our non-Hermiticities are all boundary-related and are emerging, as opposed to the ones in the new area of non-Hermitian Quantum Mechanics that are preexisting and that all seem to be of a bulk-type.

## II. EHRENFEST THEOREM

The total time derivative of the mean value of any operator that depends on position or momentum operator and has explicit time-dependence $\vec{B}(\vec{r}, \vec{p}, t)$ can be written as:

$$\frac{d}{dt}\langle \vec{B}(\vec{r},\vec{p},t)\rangle = \frac{d}{dt}\langle \Psi | \vec{B} | \Psi \rangle = \left\langle \frac{\partial \Psi}{\partial t} \Big| \vec{B} \Big| \Psi \right\rangle + \left\langle \Psi \Big| \frac{\partial \vec{B}}{\partial t} \Big| \Psi \right\rangle + \left\langle \Psi \Big| \vec{B} \Big| \frac{\partial \Psi}{\partial t} \right\rangle$$

(1)

This leads to the well-known Ehrenfest theorem of quantum mechanics[2] (usually called like this when it is applied for $\vec{B} = \vec{p}$ (or for $B=p+eA/c$), and then it defines the 'force operator', and giving the well-known velocity operator $\vec{v} = i[H,\vec{r}]/\hbar$ when it is applied for $\vec{B} = \vec{r}$). Making use of the *t*-dependent Schrodinger equation we may write

$$\left| \frac{\partial \Psi}{\partial t} \right\rangle = -\frac{i}{\hbar} H | \Psi \rangle \text{ and } \left\langle \frac{\partial \Psi}{\partial t} \right| = \frac{i}{\hbar} \langle \Psi | H^\dagger$$

for its complex conjugate. Substituting these into (1) we have

$$\frac{d}{dt}\langle \vec{B}(\vec{r},\vec{p},t)\rangle = \langle\Psi|\frac{\partial \vec{B}}{\partial t}|\Psi\rangle + \frac{i}{\hbar}\langle H\Psi|\vec{B}|\Psi\rangle - \frac{i}{\hbar}\langle\Psi|\vec{B}H|\Psi\rangle$$

$$= \langle\Psi|\frac{\partial \vec{B}}{\partial t}|\Psi\rangle + \frac{i}{\hbar}\langle H\Psi|\vec{B}|\Psi\rangle - \frac{i}{\hbar}\langle\Psi|H\vec{B}|\Psi\rangle + \frac{i}{\hbar}\langle\Psi|[H,\vec{B}]|\Psi\rangle$$

(2)

Now, if $H$ were Hermitian (with respect to $\Psi$ and $\vec{B}\Psi$), we clearly see that the result would be the familiar $\frac{d}{dt}\langle \vec{B}(\vec{r},\vec{p},t)\rangle = \langle\Psi|\frac{\partial \vec{B}}{\partial t}|\Psi\rangle + \frac{i}{\hbar}\langle\Psi|[H,\vec{B}]|\Psi\rangle$. In the more general case, however, we can rewrite (2) as:

$$\frac{d}{dt}\langle \vec{B}(\vec{r},\vec{p},t)\rangle = \langle\Psi|\frac{\partial \vec{B}}{\partial t}|\Psi\rangle + \frac{i}{\hbar}\langle [H,\vec{B}]\rangle + \frac{i}{2m\hbar}\left[\langle\Pi^2\Psi|\vec{B}|\Psi\rangle - \langle\Psi|\Pi^2\vec{B}|\Psi\rangle\right]$$

, (3)

with $\vec{\Pi}$ the kinematic momentum: $\vec{\Pi} = \vec{p} + e\vec{A}(\vec{r})/c$, with $\vec{A}(\vec{r})$ the vector potential, minimally substituted in $H$, and $\Pi^2 = p^2 - i\hbar e\vec{\nabla}.\vec{A}/c + 2e\vec{A}.\vec{p}/c + e^2 A^2/c^2$.

Substituting into (3) we get:

$$\frac{d}{dt}\langle \vec{B}(\vec{r},\vec{p},t)\rangle = \langle\Psi|\frac{\partial \vec{B}}{\partial t}|\Psi\rangle + \frac{i}{\hbar}\langle [H,\vec{B}]\rangle$$
$$- \frac{i\hbar}{2m}\iiint d^3r\left[\nabla^2\Psi^*\vec{B}\Psi - \Psi^*\nabla^2(\vec{B}\Psi)\right] \quad (4)$$
$$- \frac{e}{mc}\int d^3r\left[(\vec{\nabla}.\vec{A})\Psi^*\vec{B}\Psi + \vec{A}.\vec{\nabla}(\Psi^*\vec{B}\Psi)\right]$$

For a specific component of the vector operator $B_l(\vec{r},\vec{p},t)$ the above equation reads:

$$\frac{d}{dt}\langle B_l(\vec{r},\vec{p},t)\rangle = \langle\Psi|\frac{\partial B_l}{\partial t}|\Psi\rangle + \frac{i}{\hbar}\langle [H,B_l]\rangle - \oiint d\vec{S}.\vec{J}_{gen} \quad (5),$$

where the two volume integrals in (4) can be written as closed surface integrals (divergence theorem) on the boundary of a generalized current defined as:

$$\vec{J}_{gen} = \frac{i\hbar}{2m}\left[\vec{\nabla}\Psi^* B_l\Psi - \Psi^*\vec{\nabla}(B_l\Psi)\right] + \frac{e}{mc}\vec{A}\Psi^* B_l\Psi \quad (6)$$

This current has a form very similar to the familiar quantum probability current,

$$\vec{J}_{prob} = \frac{i\hbar}{2m}\left[\Psi\vec{\nabla}\Psi^* - \Psi^*\vec{\nabla}\Psi\right] + \frac{e}{mc}\vec{A}|\Psi|^2, \quad (7)$$

which would correspond to the special case of $B_l = 1$ (identity operator), and obeys the standard continuity equation: $\vec{\nabla}.\vec{J}_{prob} + \partial p/\partial t = 0$ with $p$ the probability density, $p = \Psi^*\Psi$. In the more general case, for any $B_l$ it can be proved that the above generalized current $\vec{J}_{gen}$ obeys a generalized continuity equation, that is violated by a nonvanishing inhomogeneous (sink) term, namely

$$\vec{\nabla}.\vec{J}_{gen} + \frac{\partial p_{gen}}{\partial t} = \Psi^*\left[\frac{\partial B_l}{\partial t} + \frac{i}{\hbar}[H,B_l]\right]\Psi, \quad (8)$$

with $p_{gen} = \Psi^* B_l \Psi$ a generalized density. To prove this, we consider the integral form of eq. (8) which is eq. (5), and upon integration in a specific volume of all terms we get:

$$\frac{d}{dt}\langle B_l(\vec{r},\vec{p},t)\rangle = \langle\Psi|\frac{\partial B_l}{\partial t}|\Psi\rangle + \frac{i}{\hbar}\langle[H,B_l]\rangle - \oiint d\vec{S}.\vec{J}_{gen} \Rightarrow$$
$$\int d^3r\frac{\partial}{\partial t}(\Psi^* B_l\Psi) = \int d^3r\Psi^*\frac{\partial B_l}{\partial t}\Psi + \frac{i}{\hbar}\int d^3r\Psi^*[H,B_l]\Psi - \int d^3r\vec{\nabla}.\vec{J}_{gen}$$

(9)

If this equality is true for any volume then we recover the differential form of generalized continuity equation, that is exactly eq. (8). Note here that, if $\partial B_l/\partial t = 0$ and if $p_{gen}$ is time independent, i.e. $\Psi$ is a single $H$-eigenstate, we have:

$$\vec{\nabla}.\vec{J}_{gen} = \frac{i}{\hbar}\Psi^*[H,B_l]\Psi \Rightarrow \oiint d\vec{S}.\vec{J}_{gen} = \frac{i}{\hbar}\langle[H,B_l]\rangle, \quad (10)$$

and $d\langle\vec{B}\rangle/dt = 0$. This means that the time derivative of mean value of any time independent operator calculated in a single stationary state is always zero.

A bit more generally, if $B_l$ is an invariant operator, $\frac{\partial B_l}{\partial t} = -i[H,B_l]/\hbar$ then $\vec{\nabla}.\vec{J}_{gen} + \partial\rho/\partial t = 0$. This is the Liouville equation. It describes the flow of $\langle B_l(\vec{r},\vec{p},t)\rangle$ through the surface (boundary of volume $V$ where the system is considered). If $B_l(\vec{r},\vec{p},t)$ is a conserved quantity, then the source term $\Sigma = \Psi^*\left[\frac{\partial B_l}{\partial t} + \frac{i}{\hbar}[H,B_l]\right]\Psi$ is zero, meaning that

$$\frac{\partial B_l}{\partial t} + \frac{i}{\hbar}[H,B_l] = 0 \Rightarrow \frac{\partial B_l}{\partial t} = -\frac{i}{\hbar}[H,B_l], \quad (11)$$

i.e $B_l(\vec{r},\vec{p},t)$ must be an invariant operator[3]. On the other hand, if the source term is nonzero, $\Sigma \neq 0$, then the above continuity equation describes the rate of flow $\Sigma \neq 0$ of the quantity $\langle B_l(\vec{r},\vec{p},t)\rangle$ in the interior of the volume $V$.

### III. HELLMANN-FEYNMAN THEOREM

Eq. (5) can be further modified if operator $B_l$ acts in a parameter space [4] as a i.e. differential operator. If we assign $B_l$ with the operator $\vec{\nabla}_{\vec{R}}$ that acts in parameter space $\{R_1, R_2, ...\}$, we get the Hellmann-Feynman theorem in a boundary-related generalized form:

$$\frac{d}{dt}\langle \vec{\nabla}_{\vec{R}} \rangle = -\frac{i}{\hbar}\langle \vec{\nabla}_{\vec{R}} H \rangle - \oiint d\vec{S} \cdot \vec{J}_{gen} , \quad (12)$$

because $\left[ H, \vec{\nabla}_{\vec{R}} \right] = -\vec{\nabla}_{\vec{R}} H$. And if we consider only one eigenstate, $\Psi = e^{-\frac{iEt}{\hbar}}|n\rangle$, we have $d\langle \vec{\nabla}_{\vec{R}} \rangle / dt = -i\vec{\nabla}_{\vec{R}} E / \hbar$ and the Hellmann-Feynman theorem (eq. (12)) becomes:

$$\vec{\nabla}_{\vec{R}} E = \langle \vec{\nabla}_{\vec{R}} H \rangle - i\hbar \oiint d\vec{S} \cdot \vec{J}_{gen} , \quad (13)$$

with $\vec{J}_{gen} = \frac{i\hbar}{2m}\left[ \vec{\nabla}\Psi^* \vec{\nabla}_{\vec{R}} \Psi - \Psi^* \vec{\nabla}(\vec{\nabla}_{\vec{R}} \Psi) \right] + \frac{e}{mc}\vec{A}\Psi^* \vec{\nabla}_{\vec{R}} \Psi$.

A rigorous Mathematical Physics presentation (through discussion of domains of definitions of operators etc.) of this type of extra boundary contributions that can show up in the Hellmann-Feynman theorem has been given in ref. [5].

## IV. EXAMPLES: (A) FREE PARTICLE

Although it is rarely mentioned, one of the main consequences of the non-Hermitian boundary terms appears already in the simplest problem of quantum mechanics: the free particle (in a volume $V$ with the standard periodic boundary conditions) whose Hamiltonian is $H = p^2/2m$ and eigenfunctions: $\Psi(\vec{r}) = e^{i\vec{k}\cdot\vec{r}}/\sqrt{V}$ (box normalization). If we choose operator $\vec{B}(\vec{r}, \vec{p}, t)$ to be the position operator, $\vec{B}(\vec{r}, \vec{p}, t) = \vec{r}$, which is clearly time independent, eq. (5) gives:

$$\frac{d}{dt}\langle x \rangle = \frac{i}{\hbar}\langle [H, x] \rangle - \frac{i\hbar}{2m}\oiint d\vec{S} \cdot \left[ \vec{\nabla}\Psi^* x\Psi - \Psi^* \vec{\nabla}(x\Psi) \right],$$

with $\langle [H, x] \rangle = -i\hbar \langle p_x \rangle / m = \hbar k_x / m$, and the second term must be evaluated on the surfaces of the cube:

$$\vec{\nabla}\Psi^* x\Psi - \Psi^* \vec{\nabla}(x\Psi) = \frac{-2i\vec{k}x + \hat{i}}{V} \Rightarrow \frac{1}{V}\oiint d\vec{S} \cdot \left[ -2i\vec{k}x + \hat{i} \right]$$
$$= \frac{1}{V}\oiint d\vec{S} \cdot \left[ (-2ik_x x + 1)\hat{i} - 2ik_y x\hat{j} - 2ik_z x\hat{k} \right] = -2ik_x \quad (14)$$

which all together result in: $\frac{d}{dt}\langle x \rangle = \frac{\hbar k_x}{m} - \frac{\hbar k_x}{m} = 0$. (15)

This is true for any of the components of $\vec{r}$ (and of course only for a single eigenstate). Note that if we had neglected the surface term in eq. (5), then $d\langle x \rangle / dt$ would not be zero, violating the condition that all mean values of time independent operators calculated in a single state must also be time independent! (a paradox earlier noted in [6] and which is also essentially what has been noticed by Quantum Chemists (as a violation of the so-called Hypervirial theorem) [7]). It is also good to notice that, if we choose $\vec{B}(\vec{r}, \vec{p}, t) = \vec{p}$ the result is once again $d\langle \vec{p} \rangle / dt = 0$ but without the appearance of a non-Hermitian boundary term (here the reason being that the momentum operator is a good self-adjoint operator for these boundary conditions).

## (B) GENERAL EXAMPLE FOR ANY GAUGE POTENTIAL: AHARONOV-BOHM CONFIGURATIONS

The fact that any mean value of a time-independent operator must not depend on time, can be generally proved for any real gauge (and vector) potential. Here we first consider for simplicity the case $\vec{A} = 0$, and examine the position operator in 1D (our method is valid for any time-independent operator, either differential or of other form)

$$\frac{d}{dt}\langle x \rangle = \frac{i}{\hbar}\langle [H, x] \rangle - \frac{i\hbar}{2m}\left[ \Phi \frac{d\Psi^*}{dx} - \Psi^* \frac{d\Phi}{dx} \right]_0^L$$
$$= \frac{\langle p \rangle}{m} - \frac{i\hbar}{2m}\left[ x\Psi \frac{d\Psi^*}{dx} - \Psi^* \frac{d(x\Psi)}{dx} \right]_0^L = \frac{\langle p \rangle}{m} - \frac{i\hbar}{2m}\left[ x\Psi \frac{d\Psi^*}{dx} - |\Psi|^2 - x\Psi^* \frac{d\Psi}{dx} \right]_0^L$$
(16)

Now, $\langle p \rangle = -i\hbar \int_0^L dx \Psi^* \frac{d\Psi}{dx}$ and by using integration by parts we conclude to:

$$\langle p \rangle = -\frac{i\hbar}{2}\left[ \left[ |\Psi|^2 \right]_0^L + \int_0^L dx \Psi^* \frac{d\Psi}{dx} - \int_0^L dx \Psi \frac{d\Psi^*}{dx} \right] \quad (17)$$

Use again integration by parts to get the second derivative of $\Psi$ with respect to x:

$$\langle p \rangle = -\frac{i\hbar}{2}\left[ |\Psi|^2 \right]_0^L - \frac{i\hbar}{2}x\left( \Psi^* \frac{d\Psi}{dx} - \Psi \frac{d\Psi^*}{dx} \right)$$
$$+ \frac{i\hbar}{2}\int_0^L dx x \frac{d}{dx}\left( \Psi^* \frac{d\Psi}{dx} \right) - \frac{i\hbar}{2}\int_0^L dx x \frac{d}{dx}\left( \Psi \frac{d\Psi^*}{dx} \right)$$
$$= -\frac{i\hbar}{2}\left[ \left[ |\Psi|^2 \right]_0^L + \left[ x\left( \Psi^* \frac{d\Psi}{dx} - \Psi \frac{d\Psi^*}{dx} \right) \right]_0^L + \int_0^L dx x \left( \Psi \frac{d^2\Psi^*}{dx^2} - \Psi^* \frac{d^2\Psi}{dx^2} \right) \right]$$
(18)

Combining (16) and (18) we find that:

$$\frac{d}{dt}\langle x \rangle = -\frac{i\hbar}{2m}\int_0^L dx x \left( \Psi \frac{d^2\Psi^*}{dx^2} - \Psi^* \frac{d^2\Psi}{dx^2} \right) \quad (19)$$

Making use of the Schrodinger equation (for a real scalar potential) we can eliminate $\Psi''$ in (19):

$$\frac{d^2\Psi}{dx^2} = -\frac{2m}{\hbar^2}[E - V(x)]\Psi,$$

to get:

$$\frac{d}{dt}\langle x \rangle = -\frac{i}{\hbar}\int_0^L dx\, x\left(-\Psi[E - V(x)]\Psi^* + \Psi^*[E - V(x)]\Psi\right) = 0$$
**(20)**

It should be noted that the above shows the necessity of including the non-Hermitian boundary terms in the case of a ring threaded by a static magnetic flux (i.e. an Aharonov-Bohm configuration [8]), so that the theorem is valid. This is in contrast to the standard literature on Aharonov-Bohm rings, where it has been stated (i.e. see [9] for a driven ring), that the Ehrenfest theorem is not valid in multiply-connected spaces.

The restoration of the above paradox can therefore also be seen as a re-establishment of the "practical applicability" of the Ehrenfest theorem in multiply-connected space.

By following the above, the reader can actually find the exact form of the non-Hermitian boundary term (or more generally of the above discussed generalized current) that heals the Ehrenfest theorem in the case of an Aharonov-Bohm ring (or, further, whenever the magnetic flux is even a time-dependent quantity). It must be noted that this non-Hermitian boundary term depends explicitly on the enclosed flux (its value, therefore, in the absence of the flux being different compared to that in the presence of a flux) – hence giving an alternative understanding of the robustness of the Aharonov-Bohm effect (and the well-known fact that the flux "cannot be gauged way").

## (C) A NOTE ON HELLMANN-FEYNMAN THEOREM IN THE BLOCH PROBLEM

Up to now, by dropping the above mentioned boundary (surface) terms, Hellmann-Feynman theorem (for differentiations with respect to a static parameter $k$) had to be written in the form[11]:

$$\frac{dE}{dk} = \left\langle \frac{dH}{dk} \right\rangle \quad \textbf{(21)}$$

Notice however that, by taking as example a Bloch electron, whose Hamiltonian (in 1D) is: $H = p^2/2m + V(x)$, with eigenfunctions $\Psi = e^{ikx}u_k(x)$, $k$ is the crystal momentum and $u_k$ is the periodic cell function, (21) results in $dE/dk = 0$, namely, it predicts that the energy bands must not depend on crystal momentum $k$. This contradicts the fact that if one minimally substitutes crystal momentum $k$ in the Hamiltonian,

$$H = \frac{(p + \hbar k)^2}{2m} + V(x), \text{ with eigenfunctions } \Psi = u_k(x),$$

eq. (21) gives

$$dE/dk = \hbar\langle(p + \hbar k)/m\rangle_u = \hbar\langle p\rangle_u/m + \hbar^2 k/m \neq 0,$$

the slope of the energy bands in a crystal. This happens because the non-Hermitian boundary term in eq. (13) coincides with zero, as we now show

$$-i\hbar\oiint d\vec{S}\cdot\vec{J}_{gen} = \frac{\hbar^2}{2m}\left[\frac{\partial u^*}{\partial x}\frac{\partial u}{\partial k} - u^*\frac{\partial}{\partial x}\left(\frac{\partial u}{\partial k}\right) + 2iku^*\frac{\partial u}{\partial k}\right]_0^L$$

With $A = -c\hbar k/e$. The above result is exactly zero, as each one of the terms appearing in is itself equal to zero, because $u_k(x = 0) = u_k(x = L)$.

What is really happening here is that $dE/dk$ is always non zero, and can be analytically obtained in full generality without the need of minimal substitution, by directly using a modified Hellmann-Feynman theorem containing our non-Hermitian boundary terms (eq. (13)):

$$\frac{dE}{dk} = \left\langle \frac{dH}{dk} \right\rangle - \frac{\hbar^2}{2m}\left[\Psi^*\frac{d}{dx}\left(\frac{d\Psi}{dk}\right) - \frac{d\Psi}{dk}\frac{d\Psi^*}{dx}\right]_0^L \quad \textbf{(22)}$$

with $H = \frac{p^2}{2m} + V(x)$ and $\Psi = e^{ikx}u_k(x)$ we have:

$$\Psi^*\frac{d}{dx}\left(\frac{d\Psi}{dk}\right) - \frac{d\Psi}{dk}\frac{d\Psi^*}{dx} = |u|^2(i - 2xk) + ix(u^*u' - uu'^*)$$
$$+ 2iku^*\frac{du}{dk} + u^*\frac{d^2u}{dxdk} - \frac{du}{dk}\frac{du^*}{dx}$$

and $dH/dk = 0$ as it should be! Substituting then in eq. (22) we obtain:

$$\frac{dE}{dk} = -\frac{\hbar^2}{2m}\left[-2|u|^2 xk + ix(u^*u' - uu'^*)\right]_0^L \quad \textbf{(23)}$$

Now, making use of eq. (18) and Schrodinger's equation:

$$u'' = -\frac{2m}{\hbar^2}\left(E - V(x) - \frac{\hbar^2 k^2}{2m}\right)u - i2ku' \text{ and}$$

$$u^{*''} = -\frac{2m}{\hbar^2}\left(E - V(x) - \frac{\hbar^2 k^2}{2m}\right)u^* + i2ku'^*,$$

we arrive at the correct result:

$$\frac{dE}{dk} = \left[\frac{\hbar^2 kx}{m}|u|^2 - \frac{i\hbar^2}{2m}x(u^*u' - uu'^*)\right]_0^L = \frac{\hbar\langle p\rangle_u}{m} + \frac{\hbar^2 k}{m} = \frac{\hbar\langle p\rangle_\Psi}{m}$$
**(24)**

that shows the consistency of the utilization of the "hidden non-Hermiticities", discussed in the present paper, with previously established results.

## (D) LINEAR COMBINATION OF STATES

$d\langle x\rangle/dt$ may not be zero only in the case of a linear combination of states as it can be easily proved using eq. (19):

$$\frac{d}{dt}\langle x\rangle = -\frac{i\hbar}{2m}\int_0^L dx\, x\left(\Psi\frac{d^2\Psi^*}{dx^2} - \Psi^*\frac{d^2\Psi}{dx^2}\right) \quad (25)$$

If $\Psi$ is a single eigenstate, then eq. (25) is zero, as shown before. But if now $\Psi$ is a linear combination of states, i.e.

$$\Psi = \sum_n C_n e^{\frac{-iE_n t}{\hbar}}\Phi_n(x), \text{ then eq. (25) becomes:}$$

$$\frac{d}{dt}\langle x\rangle = \frac{i}{\hbar}\sum_{n,l} C_l^* C_n e^{i\frac{(E_l-E_n)t}{\hbar}}(E_n - E_l)\int_0^L dx\, x\Phi_l^*(x)\Phi_n(x), \quad (26)$$

which is the correct result we obtain using elementary quantum mechanical methods. For example, consider the simple case of a particle in a quantum well (QW), with wavefunction

$\Phi_n(x) = \sqrt{\frac{2}{L}}\sin\frac{n\pi x}{L}$ with $L$ the length of QW and $n=1,2,3..$, Eq. (26) then gives:

$$\frac{d}{dt}\langle x\rangle = \frac{4i\hbar\pi^2}{mL}\sum_{n,l} C_l^* C_n e^{i\frac{\hbar\pi^2}{2mL^2}(l^2-n^2)t}\left[\frac{nl}{l^2-n^2}\right], \quad (27)$$

with the constraint: $l-n=odd$. In this case, the extra boundary contribution is still important, and some nice closed patterns can be written, but it is a matter that we currently leave to the reader.

## V. CONTRIBUTIONS OF BOUNDARY TERMS TO EHRENFEST AND HELLMANN-FEYNMAN THEOREMS WHEN THE PARAMETER HAS EXPLICIT TIME DEPENDENCE

It is interesting to recall how the Hellmann-Feynman theorem is further modified when parameters depend on time [13]. Additionally to the implicit time-dependence through the parameters, we also let $H$ depend explicitly on time $t$, i.e. $H = H(\vec{R}(t),t)$. Starting with eq. (12):

$$\frac{d}{dt}\langle\vec{\nabla}_{\vec{R}}\rangle = -\frac{i}{\hbar}\langle\vec{\nabla}_{\vec{R}} H\rangle - \oiint d\vec{S}\cdot\vec{J}_{gen}$$

with now $\frac{d}{dt} = \frac{\partial}{\partial t} + \dot{\vec{R}}\cdot\vec{\nabla}_{\vec{R}}$, the material derivative, which takes into account changes with respect to time-dependent parameters, we have:

$$\frac{d}{dt}\langle\vec{\nabla}_{\vec{R}}\rangle = -i\frac{d}{dt}\vec{A}_B = -i\left(\frac{\partial}{\partial t} + \dot{\vec{R}}\cdot\vec{\nabla}_{\vec{R}}\right)\vec{A}_B = -i\frac{\partial}{\partial t}\vec{A}_B - i\dot{\vec{R}}\cdot\vec{\nabla}_{\vec{R}}\vec{A}_B,$$
**(28)**

with $\vec{A}_B = i\langle\vec{\nabla}_{\vec{R}}\rangle$. Using then nabla product rules, we get:

$$\dot{\vec{R}}\cdot\vec{\nabla}_{\vec{R}}\vec{A}_B = \vec{\nabla}_{\vec{R}}\left(\dot{\vec{R}}\cdot\vec{A}_B\right) - \dot{\vec{R}}\times\vec{B}_B, \quad (29)$$

with $\vec{B}_B = \vec{\nabla}_{\vec{R}}\times\vec{A}_B$ and

$$\dot{\vec{R}}\cdot\vec{A}_B = i\langle\Psi|\dot{\vec{R}}\cdot\vec{\nabla}_{\vec{R}}|\Psi\rangle = i\langle\Psi|\frac{d}{dt} - \frac{\partial}{\partial t}|\Psi\rangle$$
$$= i\langle\Psi|\frac{d}{dt}|\Psi\rangle - V_B = \frac{E}{\hbar} - V_B$$
**(30)**

with $V_B = i\langle\Psi|\frac{\partial}{\partial t}|\Psi\rangle$ and $E = \langle\Psi|H|\Psi\rangle$. Combine (28), (29) and (30) to arrive at the result:

$$\langle\vec{\nabla}_{\vec{R}} H\rangle = \vec{\nabla}_{\vec{R}} E - \hbar\vec{\Omega} - \hbar\dot{\vec{R}}\times\vec{B}_B + i\hbar\oiint d\vec{S}\cdot\vec{J}_{gen}, \quad (31)$$

with $\vec{\Omega}(\vec{R},t) = -\vec{\nabla}_{\vec{R}} V_B - \frac{\partial\vec{A}_B}{\partial t}$ the "Berry electric field" and $\vec{B}_B = \vec{\nabla}_{\vec{R}}\times\vec{A}_B$ the Berry curvature, defined through potentials: $\vec{A}_B$ is the Berry vector potential (the well-known Berry connection) and $V_B$ is a "Berry scalar potential". It is interesting that equation (31) can be interpreted as describing the Lorentz force (in parameter-space) acting on a particle of charge $-\hbar$ which moves in the presence of scalar potentials $E$ and $V_B$, and a vector potential $\vec{A}_B$ (although the contribution of non-Hermitian boundary terms is generally still present and of separate importance). All quantities are defined through the full time dependent wavefunction, while, in the adiabatic limit $\dot{\vec{R}} \to 0$, they reduce to $\vec{A}_B = i\langle\Psi|\vec{\nabla}_{\vec{R}}|\Psi\rangle = i\langle n|\vec{\nabla}_{\vec{R}}|n\rangle$ and $V_B = 0$ (the standard quantities in Berry's seminal paper [11]). In the general dynamic case, the above "emergent Electromagnetism" (which, incidentally, can also incorporate Dirac "magnetic" monopoles (always in parameter-space) associated to the singularities of the Berry curvature) is expected to demonstrate a wealth of behaviors; in particular "Berry tangles" may be expected, by analogy to other areas with real magnetic fields with nonzero Gauss linking number or "magnetic helicity" [12]. This is a study that we are planning to undertake, with an eye of possible connection of the non-Hermitian boundary contributions presented in this paper to the well known bulk-boundary correspondence in topologically-nontrivial systems; by way of an example (that may have wide implications), application of this boundary term in spin-orbit coupling problems (to be presented in detail in ref.

[13]) seems to show that these non-Hermitian boundary-contributions can play a crucial role on information transfer, through an interface, between a magnetic and a non-magnetic material, that have been brought to contact.